\documentclass[aps,prb,letterpaper,amsmath,amssymb,reprint]{revtex4-1}
\usepackage{graphicx}
\usepackage{microtype}
\usepackage{bm}
\usepackage{hyperref}

\begin{document}

\title{Electrical switching of antiferromagnetic Mn$_2$Au and the role of thermal activation}
\author{Markus Meinert}
\email{meinert@physik.uni-bielefeld.de}
\author{Dominik Graulich}
\author{Tristan Matalla-Wagner}
\affiliation{Center for Spinelectronic Materials and Devices, Department of Physics, Bielefeld University, D-33501 Bielefeld, Germany}

\date{\today}

\begin{abstract}
Electrical manipulation of antiferromagnets with specific symmetries offers the prospect of creating novel, antiferromagnetic spintronic devices. Such devices aim to make use of the insensitivity to external magnetic fields and the ultrafast dynamics at the picosecond timescale intrinsic to antiferromagnets. The possibility to electrically switch antiferromagnets was first predicted for Mn$_2$Au and then experimentally observed in tetragonal CuMnAs. Here, we report on the electrical switching and detection of the N\'{e}el order in epitaxial films of Mn$_2$Au. The exponential dependences of the switching amplitude on the current density and the temperature are explained by a macroscopic thermal activation model taking into account the effect of the Joule heating in Hall cross devices and we observe that the thermal activation plays a key role in the reorientation process of the N\'{e}el order. Our model analysis shows that the electrically set N\'{e}el-state is long-term stable at room temperature, paving the way for practical applications in memory devices.
\end{abstract}

\maketitle

\section{Introduction}

The N\'{e}el-order spin-orbit torque (NSOT) was proposed as an efficient mechanism to manipulate the N\'{e}el-state of an antiferromagnet \cite{Zelezny2014, Jungwirth2016, Roy2016, Gomonay2016, Gomonay2017}. It occurs in layered antiferromagnets with specific symmetries, in particular when the two staggered sublattice magnetic moments $\bm{m}_\mathrm{A}$ and $\bm{m}_\mathrm{B}$ are inversion partners. In this case, spin polarizations $\bm{p}_\mathrm{A} = - \bm{p}_\mathrm{B}$ perpendicular to the current direction occuring due to the inverse spin galvanic effect will give rise to field-like torques $\mathrm{d}\bm{m}_\mathrm{A,B} / \mathrm{d}t \propto \bm{m}_\mathrm{A,B} \times \bm{p}_\mathrm{A,B}$ on the two sublattices with opposite sign, such that the N\'{e}el vector $\bm{L} = \bm{m}_\mathrm{A} - \bm{m}_\mathrm{B}$ can be aligned perpendicular to the electrical current. \v{Z}elezn\'y et al. predicted the NSOT in Mn$_2$Au \cite{Zelezny2014}, a tetragonal material of spacegroup $I4/mmm$ with $a = 3.328\,\mathrm{\AA{}}$ and $c = 8.539\,\mathrm{\AA{}}$ \cite{Barthem2013}. Shortly after the prediction, tetragonal CuMnAs \cite{Wadley2013} which belongs to spacegroup $P4/nmm$ was found to have similar magnetic symmetry as Mn$_2$Au, i.e., that the two magnetic sublattices are connected via inversion. The electrical switching with orthogonal current pulses and readout via the anisotropic magnetoresistance (AMR) and the planar Hall effect (PHE) were demonstrated by Wadley et al. in epitaxial CuMnAs patterned into a star-like microstructure \cite{Wadley2016}. In their pioneering work, they obtained switching amplitudes on the order of 10\,m$\Omega$ in both the AMR and PHE and found a strong dependence on the pulse width and the current density used in their switching experiments. The same group showed by x-ray magnetic linear dichroism microscopy the direct correlation between antiferromagnetic domain motion and the resistance of a simplified Hall cross device after application of orthogonal current pulses \cite{Grzybowski2017}.  Ultrafast writing of CuMnAs at ps-timescale was achieved by driving currents with pulsed THz radiation, which confirms that antiferromagnets can be written significantly faster than ferromagnets, which are limited to GHz frequencies \cite{Olejnik2018}. Recently, the electrical switching of Mn$_2$Au was observed and an efficient read-out by a giant AMR of around 5\% was confirmed \cite{Bodnar2018}.

In the present experiment, we investigate the NSOT of epitaxial Mn$_2$Au films. Film stacks of MgO (001) / ZrN 3\,nm / Mn$_2$Au 25\,nm / ZrN 2\,nm were grown by dc magnetron co-sputtering and patterned into Hall cross structures. Electrical switching experiments were performed and dependences of the switching amplitude on current density, temperature, and pulse width were studied. To interpret the results, a minimal thermal activation model was developed, for which quantitative agreement with the experimental results was obtained.

\section{Sample preparation and characterization}

\begin{figure}[t]
\includegraphics[width=8.6cm]{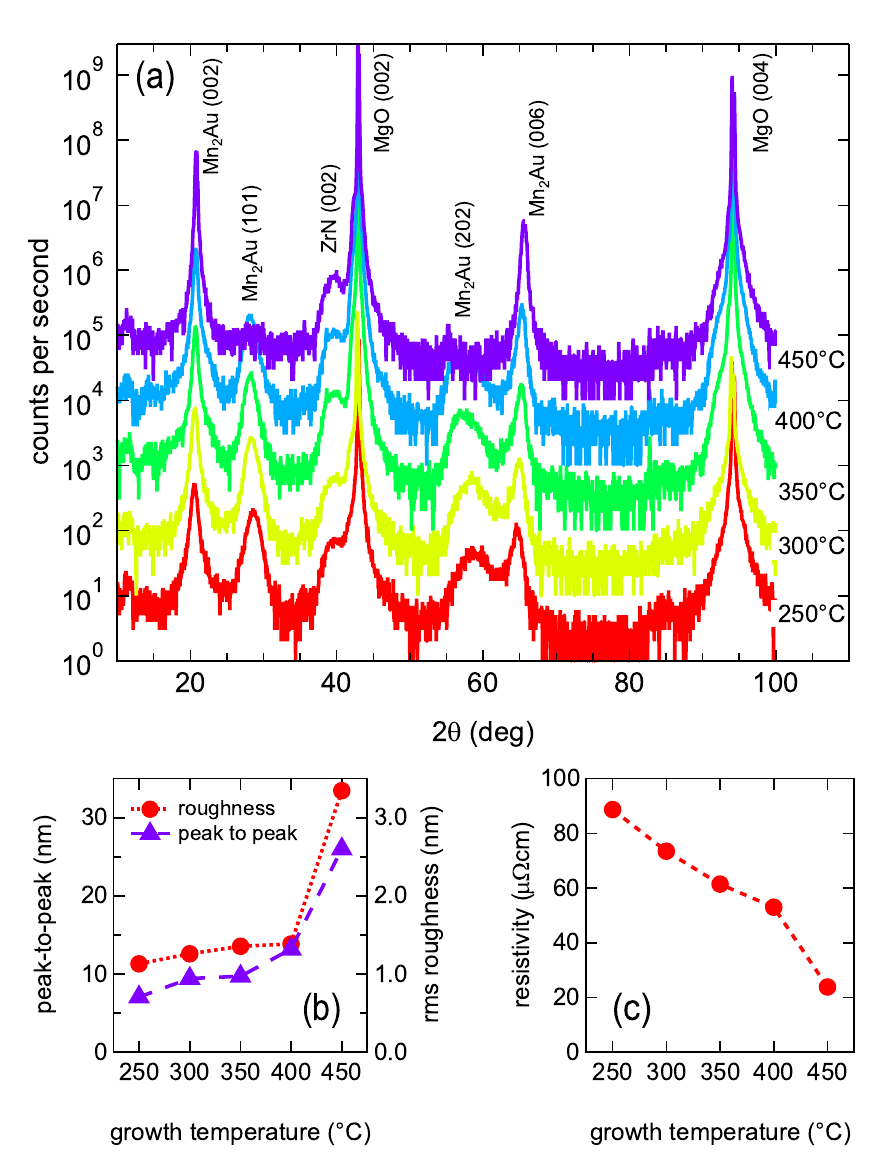}
\caption{\label{filmproperties}(a): X-ray diffraction measurements on MgO / ZrN 3 / Mn$_2$Au 25 / ZrN 2 stacks with the Mn$_2$Au layer deposited at temperatures between $250^\circ$C and $450^\circ$C. (b): Root-mean-square roughness and peak-to-peak roughness of the films obtained by AFM. (c): Resistivity of the films, obtained from four-point-measurements and assuming the nominal film thickness.}
\end{figure}

\begin{figure}[h]
\includegraphics[width=8.6cm]{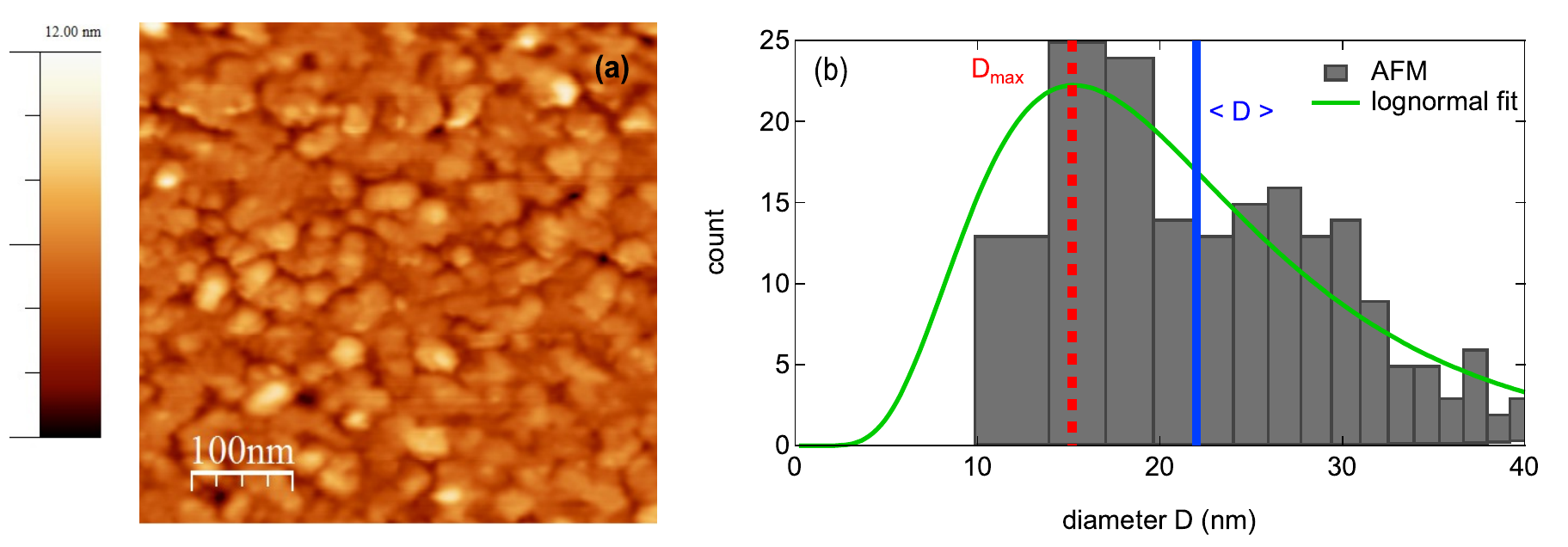}
\caption{\label{afm}Left: Atomic force microscopy image of the MgO / ZrN 3 / Mn$_2$Au 25 / ZrN 2 stack with Mn$_2$Au deposited at 300$^\circ$C used in the electrical switching study discussed in the main text. Right: Histogram of the manual grain diameter analysis of about 200 grains from the AFM picture with a lognormal fit and average/maximum probability diameters indicated by $\left<D\right>$ and $D_\mathrm{max}$, respectively. Smaller diameters than 10\,nm are not reliably obtained from the picture.}
\end{figure}

Mn$_2$Au can be grown on Al$_2$O$_3(1\bar{1}02)$ / Ta substrates \cite{Jourdan2015}, in which case we find a tilt of the $c$-axis of the Mn$_2$Au with respect to the sample surface of typically 3$^\circ$. This tilted growth can be traced back to the low-symmetry cut of the Al$_2$O$_3$ crystal. To obtain films with a preferred growth along the $c$-axis and fourfold rotational symmetry, we investigated the growth of Mn$_2$Au on MgO with various buffer layers. The thin films of Mn$_2$Au were grown by dc magnetron co-sputtering of elemental Mn and Au targets on ZrN buffered MgO substrates, resulting in stacks of the type MgO / ZrN 3 / Mn$_2$Au 25 / ZrN 2, with film thicknesses in nanometers. The ZrN buffer was deposited by reactive sputtering in an Ar/N mixture at a substrate temperature of 450$^\circ$C.  The Mn$_2$Au phase was verified by x-ray diffraction and the film composition was verified by x-ray fluorescence. The film topography was measured with an atomic force microscope (AFM). While Mn$_2$Au ($a = 3.328\,\mathrm{\AA{}}$) does not grow well on MgO ($a/\sqrt{2} = 2.977\,\mathrm{\AA{}}$) due to the large lattice mismatch of 11.5\%, it does grow on ZrN. ZrN has a lattice constant ($a/\sqrt{2} = 3.235\,\mathrm{\AA{}}$), intermediate between those of MgO and Mn$_2$Au (taking a 45$^\circ$ rotation of the basal plane into account), so that the lattice mismatch is reduced to 3\%, rendering it an ideal buffer layer giving rise to the epitaxial relation MgO [100] $\parallel$ ZrN [100] $\parallel$ Mn$_2$Au [110]. The films were finally patterned by standard optical photolithography and wire-bonded into IC packages.

In Figure \ref{filmproperties}\,(a) we show x-ray diffraction measurements of Mn$_2$Au films grown at different substrate temperatures. Pure phase Mn$_2$Au is found for all temperatures between 250$^\circ$C and 450$^\circ$C. However, the preferential orientation of the grains is improved as the temperature is increased, eventually giving pure (001)-oriented films at 450$^\circ$C. The Mn$_2$Au tends to become rough at high deposition temperatures (Fig. \ref{filmproperties}\,(b)), so that the Mn$_2$Au film used in this study was deposited at 300$^\circ$C, which results in a compromise between smoothness of the film and crystal quality. The resistivity is rather large ($73\,\mathrm{\mu \Omega cm}$, cf. Fig. \ref{filmproperties}\,(c)) and shows only weak temperature dependence (not shown), so that our constant-voltage source can be treated as a constant-current source and the current density can be simply obtained from Ohm's law with the room-temperature resistance value. In Figure \ref{afm}\,(a) we provide an AFM image of the film deposited at 300$^\circ$C. In Figure \ref{afm}\,(b), the manually evaluated grain size analysis is shown together with a lognormal fit to the histogram. The film has small grains with an average diameter of $\left< D \right> \approx 22$\,nm.

\section{Results}

\subsection{Experimental observation of switching in Mn$_2$Au}

\begin{figure*}
\includegraphics[width=\textwidth]{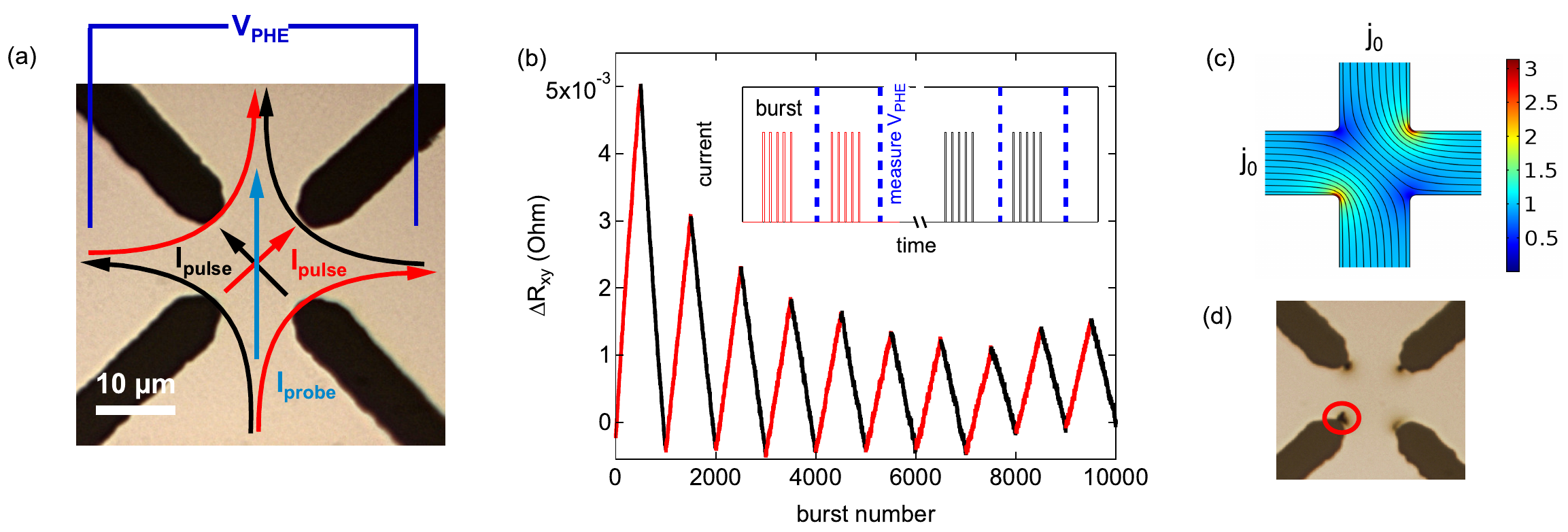}
\caption{\label{switching_demonstration}(a), Optical micrograph of the Hall cross used for the switching experiment. Current pulses are injected as indicated by red and black curved arrows, resulting in an effective current flow in the center of the Hall cross as indicated by small straight arrows. The N\'{e}el vector is thus switched at $\pm 45^\circ$ with respect to both the probe current and the voltage pickup lines, so that the planar Hall effect can be measured at maximum efficiency. (b), Demonstration of the N\'{e}el vector switching at $j_0 = 2 \times 10^{11}\,\mathrm{A/m^2}$ and $\Delta t = 200\,\mathrm{\mu s}$. Each burst consists of 100 pulses, i.e. 1.92\,mC are transferred per burst. The inset represents the pulsing and measurement scheme (not to scale). (c), Finite-element simulation of the current flow. (d), Optical micrograph of a Hall cross after heavy electrical stress. }
\end{figure*}

 For the electrical measurements, an arbitrary waveform generator (Agilent 33522A) with a differential broadband amplifier (Tabor Electronics 9260) was used as the voltage source, and a Keithley 2182A Nanovoltmeter was used to measure the planar Hall voltage $V_\mathrm{PHE}$. Simple Hall crosses were used where the current was driven through all arms simultaneously as depicted in Figure \ref{switching_demonstration}\,(a). The switching of the pulse and probe current lines was done with PhotoMOS relais. Temperature-dependent measurements were performed in a closed-cycle He cryostat. A duty cycle of 0.01 and an additional wait time of 1\,s before taking the $V_\mathrm{PHE}$ reading were found to be sufficient to remove cumulative thermal effects within the pulse bursts. Current pulse bursts with pulse widths between 100\,ns and 10\,ms were sent through the crosses and the planar Hall resistance was monitored to observe the switching as sketched in the inset of Figure \ref{switching_demonstration}(b). The bursts were programmed as to keep the transferred charge per burst constant, i.e., at larger current density $j_0$, the number of pulses per burst is reduced. As shown in Figure \ref{switching_demonstration}(b), clear piecewise linear changes of the transverse resistance $\Delta R_{xy} = V_\mathrm{PHE} / I_\mathrm{probe} -R_{xy,0}$ are seen at a current density of $j_0 = 2 \times 10^{11}\,\mathrm{A/m^2}$ and pulse width of $\Delta t = 200\,\mu$s, which changes sign when the current pulse direction is switched between two orthogonal directions, similar to the original experiment by Wadley et al. \cite{Wadley2016} A finite-element simulation of the current flow in our experiment shows that the current is very inhomogeneous and hot-spots at the corners of the Hall crosses occur (Figure \ref{switching_demonstration}(c)), which is supported by an optical micrograph of a burnt structure after applying heavy electrical stress (Figure \ref{switching_demonstration}(d)). In all measurements, we find larger switching amplitudes in the first few cycles on a given Hall cross compared to later cycles, where the amplitude remains essentially constant, cf.  Fig. \ref{switching_demonstration}(c). This is explained by the switching in the hot-spots, which see only one current direction in the experiment, so that the N\'eel vector is reoriented only once and will remain aligned for all further cycles.

\subsection{Macroscopic Model}\label{model}

A systematic investigation of the PHE resistance change (the switching amplitude) per transferred charge as functions of pulse width and current density reveals a very strong dependence of the switching amplitude on either quantity, both suggesting that thermal activation plays a crucial role in the N\'{e}el-order reorientation process. To elucidate the role of the thermal activation, we propose a macroscopic model to describe the switching.

\begin{figure}
\includegraphics[width=8.6cm]{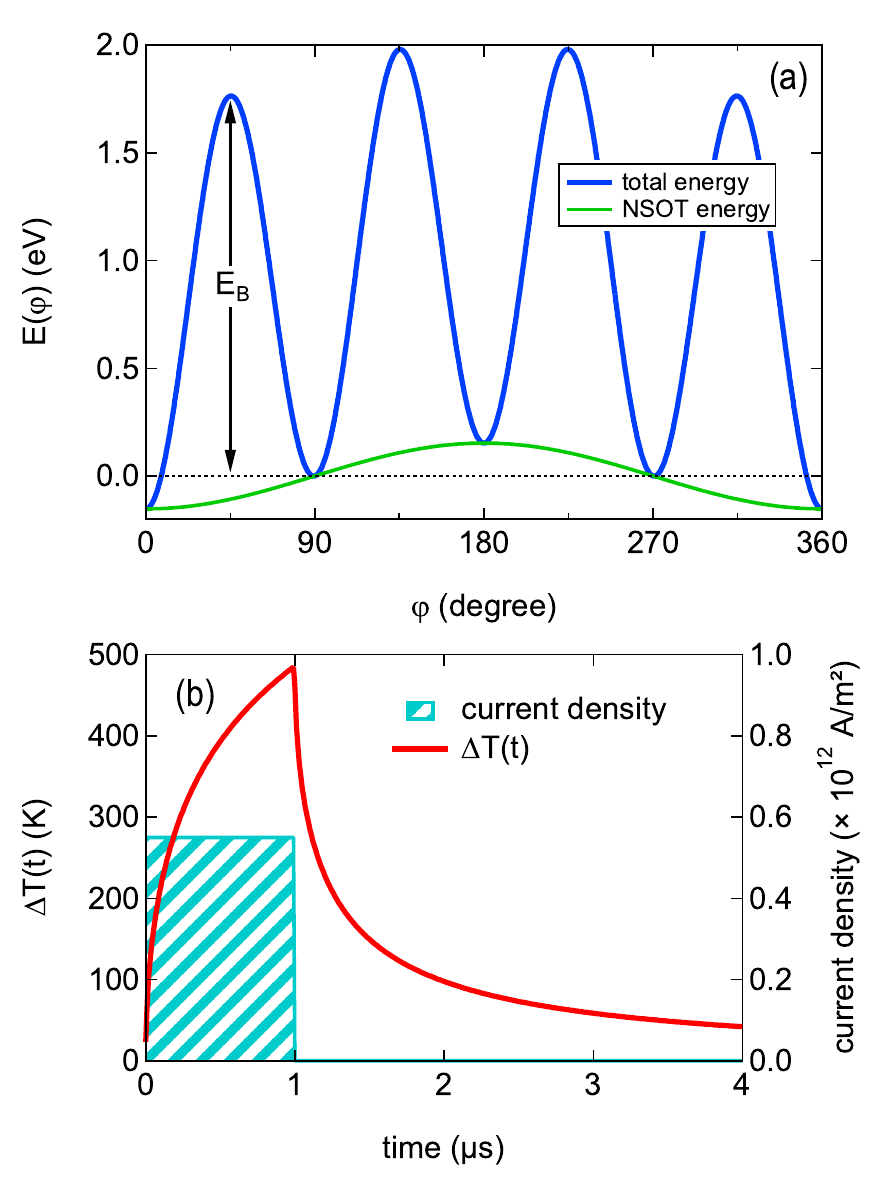}
\caption{\label{landscape_and_pulse}(a): Energy landscape $E(\varphi)$ with the parameters of Mn$_2$Au as discussed in section \ref{model}. For the NSOT energy, a current density of $j = 5.5 \times 10^{11}\,\mathrm{A/m^2}$ is assumed. The barrier $E_\mathrm{B}$ is indicated for switching from $\varphi_\mathrm{i} = 90^\circ$ to $\varphi_\mathrm{f} = 0^\circ$. b): Temperature rise $\Delta T (t)$ as described by Equation \ref{eq:T} with $1/\sigma = 73\,\mu\Omega\mathrm{cm}$, $w = \sqrt{2} \times 4\,\mu \mathrm{m}$, $\alpha = 0.5$, $j = 5.5 \times 10^{11}\,\mathrm{A/m^2}$, $h = 30\,\mathrm{nm}$ and the parameters of the MgO substrate $\kappa_\mathrm{s} = 40\,\mathrm{W/(Km)}$, $C_\mathrm{S} = 930\,\mathrm{J/(kg\,K)}$, and $\rho_\mathrm{S} = 3580\,\mathrm{kg / m^3}$. }
\end{figure}

To model the N\'{e}el-order spin-orbit torque switching, we need to describe the current-induced effective field, the energy barrier for the switching, the Joule heating of the current, and the relation between the N\'{e}el vector orientation and the planar Hall effect. The energy density of the N\'{e}el vector $\bm{L}$ in a single biaxial antiferromagnetic grain due to the staggered spin-orbit field is \cite{Roy2016, Gomonay2016}
\begin{align}\label{eq:energy}
E(\varphi) /  V_\mathrm{g} &= K_{4\parallel} \, \sin^2 2\varphi - \bm{L} \cdot \bm{B}_\mathrm{eff} \, / V_\mathrm{cell} \nonumber\\
&= K_{4\parallel} \,  \sin^2 2\varphi - |\bm{L}| |\bm{B}_\mathrm{eff}| \,  \cos\left( \varphi - \varphi_{\bm{j}} \right) / V_\mathrm{cell},
\end{align}
with $\bm{B}_\mathrm{eff} = (\bm{j} \times \bm{z}) \, \chi$ the staggered effective field, which lies in the film plane and is oriented perpendicular to the current. $K_{4\parallel}$ denotes the in-plane biaxial magnetocrystalline anisotropy energy density within the tetragonal (001) plane and $\varphi$ is the angle of $\bm{L}$ with respect to an easy axis in the (001) plane. $V_\mathrm{g}$ is the grain volume, $\bm{j}$ is the current density in the film plane, $\bm{z}$ is the unit vector along the $z$-axis perpendicular to the fim plane, $\chi$ is the spin-orbit torque efficiency given as the effective staggered field per unit current density and $V_\mathrm{cell}$ is the unit cell volume. The first term is the usual magnetocrystalline energy, whereas the second term describes the Zeeman energy of the staggered magnetic moments in the staggered spin-orbit effective field. The angular dependence of the energy is exemplarily shown in Fig. \ref{landscape_and_pulse}. The Zeeman energy stabilizes the N\'{e}el vector such that the sublattice moments will align with the staggered effective field, i.e. perpendicular to the current direction. Because of the very strong exchange interaction in Mn$_2$Au, each grain is assumed to possess collinear magnetic moments and has to rotate the N\'{e}el vector coherently (the macrospin approximation). A strong in-plane anisotropy is assumed and magnetization components pointing along the $z$-axis are neglected \cite{Barthem2013, Barthem2016}.

The energy barrier is obtained as the minimum barrier for either the clockwise (cw) or counter-clockwise (ccw) rotation from an initial state $\varphi_\mathrm{i}$ to a final state $\varphi_\mathrm{f}$:
\begin{align}\label{eq:barrier}
E_\mathrm{B} = &\min\limits_{\mathrm{cw, ccw}} \left( \max\limits_{[\varphi_\mathrm{i}, \varphi_\mathrm{f}]} \left[K_{4\parallel} \, V_\mathrm{g} \sin^2 2\varphi - \bm{L} \cdot \bm{B}_\mathrm{eff} \, V_\mathrm{g} / V_\mathrm{cell}\right] \right) \nonumber \\
&- E(\varphi_\mathrm{i}).
\end{align}

This is sketched in Figure \ref{landscape_and_pulse}\,(a) for the case of a switching event from a configuration with $\bm{L} \parallel \bm{j}$ to $\bm{L} \perp \bm{j}$. The switching rate $1/\tau$ for the switching of the N\'{e}el vector of an individual grain into a different orientation is given by the N\'{e}el-Arrhenius equation
\begin{equation}\label{eq:switching_rate}
\frac{1}{\tau} = f_0 \exp{\left( -\frac{E_\mathrm{B}}{k_\mathrm{B} T} \right)},
\end{equation}
with the attempt rate $f_0$, the Boltzmann constant $k_\mathrm{B}$ and the absolute temperature $T$. A similar approach was used to describe the dependence of the antiferromagnetic blocking temperature on the magnetocrystalline anisotropy in exchange bias systems \cite{OGrady2010}. The switching is described by the Poisson distribution, so that the switching probability with an electrical pulse of length $\Delta t$ can be written as
\begin{equation}\label{eq:switching_probability}
P_\mathrm{sw} (\Delta t) = 1 - \exp\left( - \Delta t / \tau \right).
\end{equation}

The attempt rate $f_0$ represents the picosecond dynamics of the antiferromagnet, i.e. the ultrafast precession of the sublattice magnetic moments about the exchange field \cite{Gomonay2016, Roy2016}. This description is purely kinetic and is valid for sufficiently long pulse widths $\Delta t \gg 1/f_0$, such that the dynamics of the switching can be neglected \cite{Roy2016}. The influence of the Joule heating of the current density $j = |\bm{j}|$ is taken into account by a two-dimensional model as derived by You et al.\cite{You2006} The Hall cross temperature rise as a function of time is given as
\begin{align}\label{eq:T}
\Delta T (t) &= \frac{2 w h j^2}{ \pi \kappa_\mathrm{S} \sigma } \,\mathrm{arcsinh}\left(\frac{2 \sqrt{\kappa_\mathrm{S} t / (\rho_\mathrm{S} C_\mathrm{S})}}{\alpha w}\right)\nonumber\\
&\phantom{=}+ \theta(t-\Delta t) \,\mathrm{arcsinh}\left(\frac{2 \sqrt{\kappa_\mathrm{S} (t - \Delta t) / (\rho_\mathrm{S} C_\mathrm{S})}}{\alpha w}\right).
\end{align}
Here, $h$ is the film thickness, $w$ is the current line width, $\sigma$ is the electrical conductivity, $\rho_\mathrm{S}$, $C_\mathrm{S}$, and $\kappa_\mathrm{S}$ are the mass density, specific heat, and heat conductivity of the substrate, respectively. The parameter $\alpha = 0.5$ is chosen as suggested by You et al. In Figure \ref{landscape_and_pulse}\,(b) we exemplarily show the film temperature rise for a $1\,\mu\mathrm{s}$ pulse. The temperature in Eq. \ref{eq:switching_rate} is taken as $T(t) = T_0 + \Delta T(t)$, where $T_0$ is the base temperature of the measurement. Temperature dependences of the substrate parameters are neglected. The orientation of the N\'{e}el vector is represented by a polar angle $\varphi \in [0, 2\pi)$. With this angle, the planar Hall resistance in our cross geometry is written as $R_{xy} = A \, \sin (2\varphi + \pi / 2)$ where the prefactor $A$ is equivalent to the AMR amplitude $A = |R_\parallel - R_\perp|/2$.

To find the switching current density at which the switching from an initial state into the energetically favorable orthogonal state becomes deterministic, i.e. at which the barrier becomes zero, we need to find the current density at which the local minima of the energy become saddle points. By simultaneously solving $\partial E(\varphi) / \partial\varphi = 0$ and $\partial^2 E(\varphi) / \partial\varphi^2 = 0$, we obtain 
\begin{equation}
j_\mathrm{det} = \frac{8\sqrt{6}}{9} \, \frac{K_{4\parallel}V_\mathrm{cell}}{L\chi} \approx 2.18 \, \frac{K_{4\parallel}V_\mathrm{cell}}{L\chi}.
\end{equation}

For small current density, such that $L B_\mathrm{eff} / V_\mathrm{cell} \ll K_{4\parallel}$, the local maxima of the total energy are located at $\varphi = \frac{\pi}{4}, \frac{3\pi}{4}, \frac{5\pi}{4}, \frac{7\pi}{4}$ and the barrier is simply $E_\mathrm{B} = K_{4\parallel} V_\mathrm{g} - x L B_\mathrm{eff} \, V_\mathrm{g} / V_\mathrm{cell}$, with $x = \pm 1/\sqrt{2}$, $x = \pm(1 - 1/\sqrt{2})$, or $x = -(1 + 1/\sqrt{2})$, depending on the combination of initial and final states. Only combinations with $x > 0$ are energetically favorable and facilitated by the current. The lowest barrier is found for the switching from $\bm{L} \perp \bm{B}_\mathrm{eff}$ to $\bm{L} \parallel \bm{B}_\mathrm{eff}$, i.e. $x = 1/\sqrt{2}$. Solving for the current density, we obtain
\begin{equation}
j(P_\mathrm{sw}) = \frac{V_\mathrm{cell}}{x L\chi} \left[ K_{4\parallel} - \frac{k_\mathrm{B}T}{V_\mathrm{g}} \ln\left(\frac{ - f_0 \Delta t}{\ln (1 - P_\mathrm{sw})} \right)\right].
\end{equation}

For small switching probability $P_\mathrm{sw} \approx 0$ (linear response), one can further expand the exponential and obtains $P_\mathrm{sw}(\Delta t) \approx \Delta t f_0 \exp\left(- E_\mathrm{B} / k_\mathrm{B}T\right)$. Assuming an initially uniform occupation of the four easy-axis configurations and assuming all grains are equal, the planar Hall resistance change reads $| \Delta R_{xy} | = A P_\mathrm{sw}$. Therefore, we obtain for the planar Hall resistance change per pulse in the linear response regime
\begin{equation}\label{eq:LinResponse}
\left| \Delta R_{xy} \right| \approx A \Delta t f_0 \exp\left( - \frac{K_{4\parallel}V_\mathrm{g}}{k_\mathrm{B}T} \right)\exp \left( \frac{xLj\chi V_\mathrm{g}}{k_\mathrm{B}T V_\mathrm{cell}} \right).
\end{equation}
In this regime, we expect to find a linear dependence of the switching amplitude on the pulse width. The switching amplitude is predicted to depend exponentially on the current density and also exponentially on the measurement temperature.

\begin{figure*}
\includegraphics[width=\textwidth]{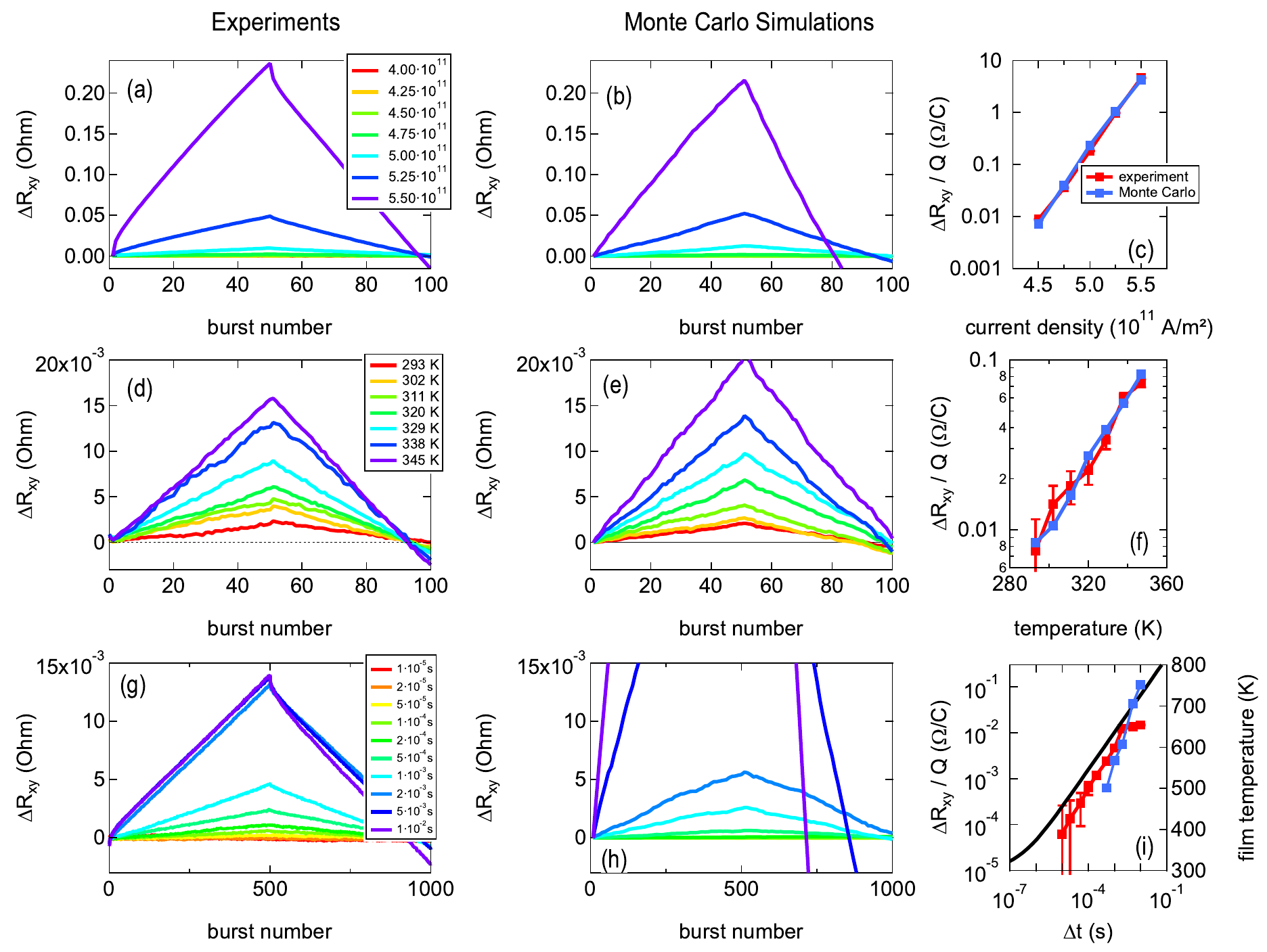}
\caption{\label{comparison_experiment_MC} (a)-(c), switching amplitude as a function of the current density $j_0$ (given as $\mathrm{A/m^2}$). Experiment with $\Delta t = 1 \times 10^{-6}\,\mathrm{s}$ and 1\,mC per burst, (a), Monte Carlo simulations with $D = 22.0\,\mathrm{nm}$ and $K_{4\parallel} = 7.5\,\mathrm{\mu eV}$ per cell, (b), and extracted switching amplitudes per unit charge (c). (d)-(f), same as (a)-(c) for variations of the base temperature with $j_0 = 4.25 \times 10^{11}\,\mathrm{A/m^2}$, $\Delta t = 1 \times 10^{-6}\,\mathrm{s}$, and 5\,mC per burst. The Monte Carlo simulations were performed with the same grain size and anisotropy energy density as in (b) and (c). (g)-(i), same as (a)-(c) for variations of the pulse width with 2\,mC per burst and $j_0 = 2 \times 10^{11}\,\mathrm{A/m^2}$. The black solid line in (i) represents the calculated film temperature according to equation \ref{eq:T}}.
\end{figure*}

To fully take into account the influence of the pulse sequence and the film temperature due to the Joule heating, numerical simulations were performed with a Monte Carlo technique. An ensemble of $N = 10^5$ grains with N\'{e}el vector orientations $\varphi$ evenly distributed among the four easy-axis configurations was set-up. Optionally, the grain sizes can be taken from any distribution, e.g., a lognormal distribution. Here, only one fixed grain diameter is considered, which is treated as a phenomenological parameter. We use the average grain diameter as obtained by AFM, i.e. $\left< D \right> \approx 22\,\mathrm{nm}$. The time-evolution of the film temperature is taken into account by a time-discretization scheme, such that multiple kinetic Monte Carlo trials are done for each pulse during heat-up (with current applied) and cool-down (without current) phases. For each trial, a random trial configuration is set up and its energy barrier according to Eq. \ref{eq:barrier} and the switching probability according to Eq. \ref{eq:switching_probability} are computed for each grain independently. Random numbers $R \in [0, 1)$ are generated for each grain and the switching into the trial state is accepted if $P_\mathrm{sw} > R$. The pulse burst patterns from the experiment are implemented in the model calculations to accurately reflect the experimental conditions and allow for a one-to-one comparison between the simulated and the experimental data sets. The model allows the N\'eel vector to switch into any easy-axis configuration, however, the probability to do so is larger if the sublattice magnetic moments are parallel to the staggered effective field. Thus, also thermally activated back-hopping is included in the model.

All quantities entering into the equations can in principle be obtained from experiments or \textit{ab initio} calculations. We take $|\bm{L}| = 2 \times 4\mu_\mathrm{B}$ and $V_\mathrm{cell} = 4.75 \times 10^{-29}\,\mathrm{m^3}$ \cite{Barthem2013}. The film had a resistivity of $73 \times 10^{-8}\, \Omega\mathrm{m}$ and the thermal parameters of the MgO substrate are taken from Ref. \cite{Stackhouse2010} (cf. Fig. \ref{landscape_and_pulse}). The attempt rate $f_0$ is taken as $f_0 = 10^{12}\,\mathrm{s}^{-1}$.\cite{Arana2017} The core region of the Hall cross has a resistance of about 20\,$\Omega$ and we estimate an AMR amplitude of 5\,\% based on our observations, so that the planar Hall effect amplitude is $A \approx 1\,\Omega$. This estimate is consistent with a recent calculation of the AMR in Mn$_2$Au \cite{Bodnar2018}. The current density at the center of the constriction of the Hall cross is used in the calculations, i.e. $j \approx j_0$, cf. Figure \ref{switching_demonstration}\,(c). As is evident from Eq. \ref{eq:LinResponse}, the products $K_{4\parallel} V_\mathrm{g}$ and $\chi V_\mathrm{g}$ can be directly obtained by fitting the measured switching amplitudes as a function of the current density. Good agreement with the experiment is obtained with $K_{4\parallel} V_\mathrm{g} \approx 1.5\,\mathrm{eV}$. The parameters $V_\mathrm{g}$, $\chi$, and $K_{4\parallel}$ are difficult to obtain individually and are treated as adjustable parameters that will be discussed in the following sections.

\subsection{Model results}

In Figures \ref{comparison_experiment_MC}(a),(d),(g) we show measurements of the switching amplitude for variations of the current density, the base temperature, and the pulse width, in a side-by-side comparison with the corresponding Monte Carlo simulations in Figures \ref{comparison_experiment_MC}(b),(e),(h) and the amplitudes per unit charge in Figures \ref{comparison_experiment_MC}(c),(f),(i). The dependences on the current density and on the temperature are very well reproduced by the model calculations with reasonable choices for the anisotropy energy density, providing strong evidence for the central role of the thermal activation in the switching process. Specifically, the clear increase of the switching amplitude by increasing the base temperature $T_0$ of the measurement (Figure \ref{comparison_experiment_MC}(d)) shows that enhanced film temperature due to Joule heating will enhance the switching amplitude. For comparison, we performed calculations of the pulse width and current density variations without including the effect of Joule heating (not shown). In that case, the dependence on the current density is much less pronounced and the dependence on the pulse width (keeping the transferred charge per burst constant) will vanish completely, in contrast to the experiment. The switching amplitude increases exponentially with current density and temperature which is easily seen in the logarithmic plots in Figures \ref{comparison_experiment_MC}(c),(f). The model reproduces these exponential dependencies quite accurately. The discrepancies between the experiments and the model calculations are somewhat larger for the pulse width dependence. As was shown by Fangohr et al., Equation \ref{eq:T} has limited applicability for short wires as in the case of the present experiment \cite{Fangohr2011}. The maximum pulse duration within which acceptable results are expected is given by $t_\mathrm{c} \approx (L/2)^2 \rho_\mathrm{S} C_\mathrm{S} / \kappa_\mathrm{S}$. In the present case, we estimate the limit to be 10 to 20\,$\mu$s, above which the film temperature increases less strongly than predicted by the formula due to the three-dimensional heat flow. Therefore, the pulse width dependence is not well reproduced by the model calculation, whereas the current density and temperature dependences taken with a pulse width of 1\,$\mu$s are very well described by the model. The saturation of the switching amplitude seen in the experimental pulse width dependence can thus be traced back to a saturation of the film temperature as a result of the three-dimensional heat flow.

\subsection{Discussion}

 Without a current applied, the energy barrier is simply $E_\mathrm{B} = K_{4\parallel} V_\mathrm{g}$. Thus, the thermal stability factor at room temperature  $\Delta = E_\mathrm{B} / k_\mathrm{B} T \approx 60$ is large enough to ensure a stable N\'{e}el-state over time scales of more than ten years, and meets an important prerequisite for permanent magnetic memories \cite{Khvalkovskiy2013}. By measuring the grain size distribution with atomic force microscopy, we found the average grain diameter $\left< D \right> \approx 22\,\mathrm{nm}$ and, correspondingly, $K_{4\parallel} \approx 7.5\,\mu \mathrm{eV} / \mathrm{f.u.}$. From a recent measurement of the antiferromagnetic zone-center magnon modes by Brillouin light scattering, Arana et al. obtained a basal-plane anisotropy field of $0.007\,\mathrm{T}$ \cite{Arana2017}, i.e. $K \approx 3.2\,\mu \mathrm{eV} / \mathrm{f.u.}$ and is in fair agreement with our estimate. However, our simulations taking the lognormal distribution as obtained from the AFM image into account disagree with the experiment: less pronounced dependences of the switching amplitude on temperature and current density are expected in that case (not shown). Assuming the anisotropy energy density obtained by Arana et al. is also valid for our samples, the corresponding grain diameter is $D \approx 35.2\,\mathrm{nm}$, which is significantly larger than most of the observed grains. Both discrepancies can be explained by assuming that antiferromagnetic domains are formed across grain boundaries, thus smaller grains are stabilized by the intergranular exchange interaction and are harder to switch. However, we note that our model does not provide an understanding of the microscopic switching behaviour and does not allow to draw detailed conclusions about the formation of antiferromagnetic domains, which have to be obtained by other means. With the present estimates for the grain or domain size, we can estimate the spin-orbit torque efficiency as $\chi = 0.25 \dots 0.12 \, \mathrm{mT /( 10^{11}\, A/m^2)}$, where the higher value corresponds to the smaller grains and agrees reasonably well with the calculated value of $\chi = 0.3 \, \mathrm{mT / (10^{11}\, A/m^2)}$ by \v{Z}elezn\'y et al. \cite{Zelezny2014} The current density for deterministic switching can be written as $j_\mathrm{det} \approx (2.18 K_{4\parallel} V_\mathrm{g}) / (\chi V_\mathrm{g}) \cdot  (V_\mathrm{cell} / L) \approx 1.4\times 10^{13}\,\mathrm{A/m^2}$. All parameters entering into this expression are accessible from the experiment and $j_\mathrm{det}$ is independent of the grain volume. Because of the strong heating that destroys the films, this curent density is experimentally only accessible with ultrafast current pulses \cite{Olejnik2018}.

\begin{figure}
\includegraphics[width=8.6cm]{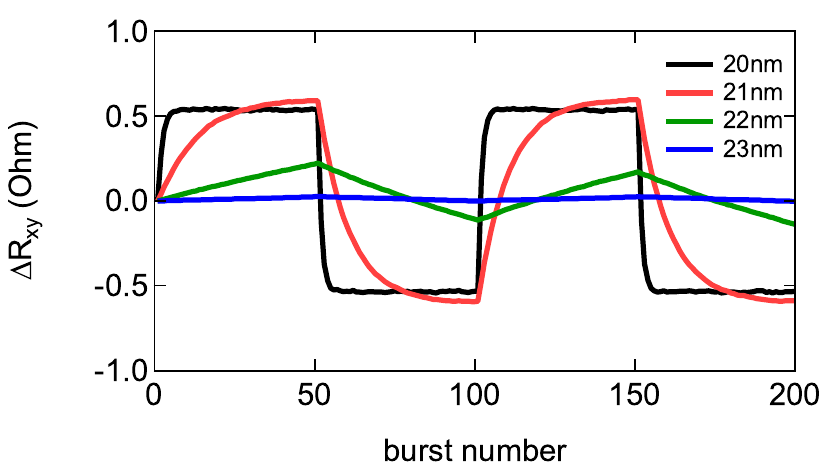}
\caption{\label{graindiameter_dependence}Calculations with different grain diameters and $\Delta t = 1\,\mathrm{\mu s}$, $j = 5.5\times 10^{11}\,\mathrm{A/m^2}$ and 1\,mC per burst at $T_0 = 293\,\mathrm{K}$.}
\end{figure}

 In Figure \ref{graindiameter_dependence} we show the strong sensitivity of the switching amplitude on the grain diameter. Increasing the grain diameter by 4\,nm, one can tune the switching behaviour from saturated to essentially not switchable at a given set of experimental conditions. The grain size and the film thickness are thus critical parameters for the NSOT switching. Due to the high film temperature, the switching amplitude is suppressed, because switching into energetically unfavorable configurations is facilitated through thermal activation. The initial kinks in the switching seen, e.g., in Figure \ref{comparison_experiment_MC}(a) are hereby explained by the presence of a small number of smaller grains that are easy to saturate, whereas the majority of the grains show the linear behaviour far off saturation. The apparently exponential saturation observed by Wadley et al. in CuMnAs\cite{Wadley2016} is explained by our model as a consequence of the stochastical nature of the switching process and the underlying Poisson probability distribution.

\section{Conclusion}
To summarize, we prepared epitaxial Mn$_2$Au films and demonstrated the electrical switching of the N\'{e}el vector. A simple expression for the switching energy barrier was proposed and implemented into a principally parameter-free Monte Carlo model, which shows quantitative agreement with the measurements with realistic model parameters. The analytic expression for the transverse resistance change provides a transparent picture of the influence of the individual experimental parameters. The Joule heating of the current plays a key role for the switching process, as it provides sufficient thermal activation to reorient the N\'{e}el vector with the weak spin-orbit torque. The switching energy barrier in Mn$_2$Au is large enough to retain the magnetic state over many years at room temperature, and is suitable for a practical application in antiferromagnetic spintronic devices. 

\begin{acknowledgments}
The authors thank G. Reiss for making available the laboratory equipment. They further thank D. Kappe for providing the finite-element simulation of the Hall cross device.
\end{acknowledgments}

\end{document}